\documentclass[reprint,superscriptaddress,amsmath,amssymb,twocolumn,prl]{revtex4-2}

\usepackage{amsmath}
\usepackage{mathtools}
\usepackage{amssymb}
\usepackage{graphicx}
\usepackage{bm}
\usepackage[colorlinks, linkcolor=blue, citecolor=blue, urlcolor=blue, breaklinks=true]{hyperref}
\usepackage{color,dsfont,multirow,booktabs}
\usepackage{braket}
\usepackage{tabularx}
\usepackage{lipsum,booktabs}
\usepackage{subfiles}
\usepackage{subcaption}
\usepackage{caption}
\usepackage[dvipsnames]{xcolor}
\usepackage{ragged2e}
\usepackage{orcidlink}
\usepackage[T1]{fontenc}
\usepackage[normalem]{ulem}

\newcommand{\commentold}[1]{}
\DeclareMathSymbol{:}{\mathpunct}{operators}{"3A}

\newcommand{\figpanel}[2]{\hyperref[#1]{\ref*{#1}(#2)}}

\begin{document}
\date{\today}

\title{Quantum-Battery-Powered Geometric Landau--Zener Interferometry}

\author{Borhan Ahmadi\orcidlink{0000-0002-2787-9321}}
\email{borhan.ahmadi@ug.edu.pl}
\address{International Centre for Theory of Quantum Technologies, University of Gdańsk, ul. prof. Marii Janion 4, 80-309 Gdańsk, Poland}

\begin{abstract}
Classical microwave drives are usually treated as ideal phase-coherent work sources for superconducting-qubit control. What if such a drive is replaced by a finite quantum battery. As a demanding benchmark, we consider echo-refocused geometric Landau--Zener interferometry powered by a single quantized bosonic mode. The qubit--battery dynamics are described by a Jaynes--Cummings Hamiltonian, while the echo pulse is retained as a qubit-only refocusing operation that cancels the dynamical phase. In the macroscopic coherent-state limit, the usual classical geometric interferometer is recovered. At finite mean photon number, however, the Jaynes--Cummings coupling generates photon-number-resolved avoided crossings with gaps \(\Omega_n=2g\sqrt{n}\). The qubit-only echo redistributes amplitudes between neighboring excitation sectors, so the finite-battery protocol is not a single classical interferometer but a coherent sector-resolved quantum evolution. This produces contrast loss, interferogram distortions, and measurable battery back-action. We further show that reducing photon-number fluctuations alone is not sufficient: geometric control requires a first-order phase reference. Geometric Landau--Zener interferometry therefore provides a practical benchmark for certifying phase-coherent quantum-battery energy.
\end{abstract}

\maketitle
\textit{Introduction}---
Coherent control in superconducting circuits is usually formulated in terms of prescribed classical microwave fields~\cite{Blais2021}. In this description, the field amplitude and phase are external resources: the source, control electronics, microwave line, attenuation chain, and heat load do not appear in the quantum Hamiltonian. In actual processors, these hidden resources become physical bottlenecks, since phase-coherent microwave signals are synthesized outside the cryostat and routed to the chip through many control channels, whose wiring density, passive heat load, and electronics overhead obstruct scaling~\cite{Bao2024}. This has motivated cryogenic control, on-chip microwave sources, and the proposal to use a precharged bosonic mode as an intrinsic quantum battery (QB) that remains coherent with the qubits it powers~\cite{PhysRevE.87.042123,PhysRevLett.118.150601,l39v-jwwz}. Such a battery would not merely deliver energy; it would participate in the coherent dynamics. The central issue is then whether phase-coherent control requires a macroscopic drive field, or can survive when the source contains only a few quanta.

Quantum batteries have so far been assessed primarily through energetic figures of merit: stored energy and extractable work~\cite{RevModPhys.96.031001,PhysRevE.87.042123,PhysRevA.107.042419}, charging power and quantum charging advantages~\cite{PRXQuantum.5.030319,PhysRevLett.118.150601,PhysRevLett.120.117702,PhysRevLett.127.100601,Rodriguez_2024,zakavati2025optimizing,PhysRevA.106.042601,whs2-dkgh}, and open-system performance under dissipative charging, steady-state operation, reservoir engineering, and noise-assisted storage~\cite{stable1,kamin2023steady,PhysRevApplied.23.024010,PhysRevE.105.064119,ahmadi2025harnessing,grazi2026charging,hadipour2025nonequilibrium}. Related work has emphasized self-discharge mitigation, degradation suppression, localization, topology, nonreciprocity, and chiral transport as routes to preserving stored energy or useful work~\cite{PhysRevResearch.2.013095,d9k1-75d4,PhysRevA.102.060201,PhysRevE.108.064106,PhysRevLett.132.090401,PhysRevLett.134.180401,PhysRevLett.132.210402,liu2026chiral}, while more recent operational approaches focus on useful charge, charge-preserving transformations, ergotropy protection, and reliable work transfer~\cite{2jtp-jpkn,bv4w-jr6q,6kwv-z6fx,9vv8-s8r1,e28040396,ahmadi2025reservoir}. These criteria are essential, but they do not by themselves certify that a battery can replace a phase-coherent control field. For coherent operations, the source must define not only an energy scale, but also a controllable transverse phase.

A stringent way to test this distinction is to ask whether a finite QB can power an interferometric control protocol. Landau--Zener--Stueckelberg interferometry turns repeated passages through an avoided crossing into the analogue of a Mach--Zehnder interferometer: each Landau--Zener transition acts as a beam splitter, while the phase accumulated between passages determines the final population~\cite{Landau1932,Zener1932,Stueckelberg1932,Shevchenko2010,Sillanpaa2006}. A geometric variant sharpens this test: by inserting an echo pulse between the two passages, the dynamical phase can be canceled, leaving a fringe controlled by the geometric phase~\cite{Gasparinetti2011,Tan2014}. The resulting protocol is therefore a demanding benchmark for phase-stable quantum control.

Several adjacent directions have been explored before. Quantized bosonic modes can modify, enable, or resolve Landau--Zener dynamics through the Jaynes--Cummings sector structure~\cite{Saito2006,Wubs2007,Keeling2008,Fink2008,Sun2012,Ashhab2014,Malla2018}, and related work has examined geometric control with quantized fields~\cite{Siddiqui2006,Zheng2012}. Recent QB proposals have also considered precharged bosonic modes as resources for powering energy-exchange gates~\cite{l39v-jwwz}. Conversely, superconducting demonstrations of Landau--Zener and geometric Landau--Zener interferometry used prescribed classical drives~\cite{Sillanpaa2006,Gasparinetti2011,Tan2014}. Here we combine these directions by using a single quantized bosonic mode to power the transverse Landau--Zener passages of geometric interferometry, and by identifying phase-coherent battery energy, not energy alone, as the resource certified by high-contrast geometric control.

We replace the prescribed transverse microwave drive by a bosonic QB and evolve the full qubit--battery system under the Jaynes--Cummings Hamiltonian~\cite{JaynesCummings1963}. The echo pulse is retained as an ideal qubit-only refocusing operation, applied identically in the classical and finite-QB benchmarks, so the protocol isolates whether a finite bosonic source can supply the phase-coherent transverse coupling that forms the Landau--Zener beam splitters. In the macroscopic coherent-state limit, the usual classical drive is recovered. At finite mean photon number, however, the Jaynes--Cummings coupling generates
a coherent bundle of photon-number-resolved Landau--Zener passages: during each
sweep segment, the pair \(|n,g\rangle\) and \(|n-1,e\rangle\) is coupled with gap
\(\Omega_n=2g\sqrt{n}=\Omega\sqrt{n/\bar n}\), while the qubit-only echo redistributes
amplitudes between neighboring excitation sectors.

This sector structure broadens the effective Landau--Zener gap and reduces the geometric-fringe contrast, but it also exposes a distinct requirement: the battery must provide a phase reference. We quantify this by the coherent fraction \(\eta_{\rm coh}=|\langle a_{\rm b}\rangle|^2/\langle a_{\rm b}^\dagger a_{\rm b}\rangle\). Number-squeezed states can narrow the gap distribution, but at fixed energy they can also reduce the coherent displacement that defines the transverse control phase. Thus the relevant resource is not stored energy alone, but phase-coherent battery energy. We show that a coherent QB approaches the classical geometric interferometer as \(\bar n\) increases, while few-quanta batteries still produce clear fringes, with a finite contrast deficit and measurable back-action. Geometric Landau--Zener interferometry therefore provides a practical benchmark for QB-powered coherent control.

\begin{figure}
    \centering
    \includegraphics[width=1\linewidth]{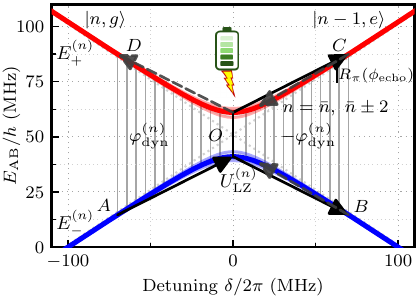}
    \caption{Photon-number-resolved Landau--Zener passages powered by a finite quantum battery. During each Jaynes--Cummings sweep segment, \(|n,g\rangle\) and \(|n-1,e\rangle\) form dressed adiabatic branches \(E_{\rm AB}^{(n)}\) separated by \(\Omega_n=\Omega\sqrt{n/\bar n}\), shown for \(n=\bar n\) and \(n=\bar n\pm2\) with \(\bar n=5\). Solid arrows indicate the forward passage. The qubit-only echo \(R_\pi(\phi_{\rm echo})\) swaps the qubit state without changing the battery photon number, so the dashed arrows are schematic refocused paths rather than exact same-\(n\) trajectories. The full finite-QB calculation includes this echo-induced redistribution between neighboring Jaynes--Cummings sectors. \justifying}\label{Sketch}
\end{figure}

Figure~\ref{Sketch} summarizes the sector-resolved Landau--Zener passage picture that schematically underlies the finite-battery dynamics. During each Jaynes--Cummings sweep segment, the QB couples \(|n,g\rangle\) to \(|n-1,e\rangle\), producing dressed adiabatic branches \(E_{\rm AB}^{(n)}\) separated by the gap \(\Omega_n=\Omega\sqrt{n/\bar n}\). The qubit-only echo swaps the qubit state without changing the battery photon number, and therefore maps amplitudes into neighboring excitation sectors before the reverse passage. Thus the figure should be read as a dressed-gap picture of the Landau--Zener passages, not as an exact independent-sector decomposition of the full echo-refocused sequence. The full simulations include this echo-induced sector redistribution. The remaining geometric fringe tests whether the finite battery supplies the azimuthal phase reference required for coherent control.

\textit{Model}.--
We consider a two-level system whose geometric Landau--Zener interference is powered by a single quantized bosonic mode. The mode acts as a finite quantum battery: it supplies the energy exchanged with the qubit and also provides the phase reference that fixes the azimuthal direction of the transverse coupling. In a rotating frame and within the rotating-wave approximation, the joint Hamiltonian is
\begin{equation}
    H(t) = \frac{\delta(t)}{2}\sigma_z + g\left(a_{\rm b}\sigma_+ + a_{\rm b}^\dagger\sigma_-\right),
\label{eq:H_QB}
\end{equation}
where $a_{\rm b}$ annihilates a battery excitation, $g$ is the qubit--battery
coupling strength, and $\sigma_+=|e\rangle\langle g|$, $\sigma_- = |g\rangle\langle e|$. The detuning $\delta(t)$ is swept through an avoided crossing twice, so that the two Landau--Zener passages act as beam splitters for the qubit amplitudes. Between the two passages we apply an echo pulse, modeled as an instantaneous $\pi$ rotation about an axis in the equatorial plane, which cancels the dynamical phase. The output of the interferometer is the final excited-state population $P_e={\rm Tr}[\rho(\tau_C)|e\rangle\langle e|]$, after tracing over the battery.

The detuning protocol consists of a forward sweep from $-\delta_0$ to $+\delta_0$ during a time $\tau_p$, a plateau at $+\delta_0$, and a reverse sweep back to $-\delta_0$ during a second time $\tau_p$:
\begin{equation}
    \delta(t) =
    \begin{cases}
    -\delta_0+2\delta_0 t/\tau_p,
    &
    0\leq t\leq \tau_p,
    \\[1mm]
    +\delta_0,
    &
    \tau_p<t<\tau_C-\tau_p,
    \\[1mm]
    +\delta_0
    -
    2\delta_0
    \left[
        t-(\tau_C-\tau_p)
    \right]/\tau_p,
    &
    \tau_C-\tau_p\leq t\leq\tau_C .
    \end{cases}
\label{eq:detuning_waveform}
\end{equation}
The echo pulse is applied at the midpoint of the plateau, \(t_m=\tau_C/2\), and is treated as an instantaneous qubit rotation,
\(
    U_\pi(\phi_{\rm echo})
    =
    \exp\left[
        -\frac{i\pi}{2}
        \left(
            \cos\phi_{\rm echo}\,\sigma_x
            +
            \sin\phi_{\rm echo}\,\sigma_y
        \right)
    \right].
\)
In the full qubit--battery Hilbert space this operation acts as
\(
    \rho(t_m^+)
    =
    \left(
        \mathbb{I}_B\otimes U_\pi(\phi_{\rm echo})
    \right)
    \rho(t_m^-)
    \left(
        \mathbb{I}_B\otimes U_\pi^\dagger(\phi_{\rm echo})
    \right),
\)
where \(\mathbb{I}_B\) is the identity on the battery mode. Thus the echo pulse does not act on the battery directly and, unlike the Jaynes--Cummings Hamiltonian, it does not conserve \(N_{\rm tot}=a_b^\dagger a_b+|e\rangle\langle e|\). For example, up to an overall phase, \(U_\pi|g\rangle=|e\rangle\) and \(U_\pi|e\rangle=|g\rangle\), so \(|n,g\rangle\) is mapped to \(|n,e\rangle\), while \(|n-1,e\rangle\) is mapped to \(|n-1,g\rangle\). These states belong to neighboring Jaynes--Cummings excitation sectors. Therefore the photon-number-resolved gap \(\Omega_n=2g\sqrt n\) is an exact block label during each sweep segment, but not a conserved label of the entire echo-refocused sequence. The role of the echo is to remove the deterministic dynamical phase in the classical reference protocol and to define the axis relative to which the battery phase is scanned; the finite-QB dynamics, including the echo-induced redistribution between sectors, is obtained from the full density-matrix evolution below. The qubit-only echo redistributes amplitudes between neighboring Jaynes--Cummings sectors; the associated gaps satisfy
\(\Omega_{n\pm1}=\Omega_n[1\pm (2n)^{-1}+O(n^{-2})]\), so this effect is asymptotically small at large \(n\) but is part of the intrinsic few-quanta finite-battery physics (see Note~1 of the SM~\cite{SuppMat}).

The geometric phase is controlled by the initial phase of the battery. For a coherent battery we take $|\psi_B(0)\rangle=|\alpha_\theta\rangle$, with $\alpha_\theta=\sqrt{\bar n}\,e^{-i\phi_\theta}$ and $\phi_\theta=\theta_{\rm geo}-\pi/2$. Choosing $g=\Omega/(2\sqrt{\bar n})$ keeps the mean transverse gap fixed as $\bar n$ is varied. In the mean-field limit, replacing $a_{\rm b}$ by $\langle a_{\rm b}\rangle$ gives $g(\langle a_{\rm b}\rangle\sigma_+ + \langle a_{\rm b}^\dagger\rangle\sigma_-) = (\Omega/2)(\cos\phi_\theta\,\sigma_x+\sin\phi_\theta\,\sigma_y)$. Thus the usual classical transverse drive is recovered as the large-$\bar n$ limit of the quantized battery model, while finite $\bar n$ exposes corrections that are invisible when the drive is treated as an ideal external field.

The photon-number-resolved gaps \(\Omega_n=2g\sqrt n\) provide the microscopic origin of the finite-battery broadening discussed below and derived in Note~1 of the SM~\cite{SuppMat}. A battery with photon-number distribution \(p_n\) therefore generates a coherent superposition of Landau--Zener passages with different gaps. Because the qubit-only echo changes the excitation-sector label, these passages should not be interpreted as independent closed interferometers in fixed \(n\)-blocks; rather, they provide the sector-resolved dressed-gap mechanism that controls the full finite-QB evolution. For $n=\bar n+\delta n$ and
$|\delta n|\ll\bar n$, the relative gap fluctuation is
$\delta\Omega_n/\Omega\simeq \delta n/(2\bar n)$, and hence
$\Delta\Omega/\Omega\simeq \sqrt{{\rm Var}(n)}/(2\bar n)$ (see Note~1 of the SM~\cite{SuppMat}). Photon number fluctuations therefore produce an inhomogeneous broadening of the Landau--Zener gap and reduce the interference contrast.

To include relaxation, dephasing, and battery loss, we evolve the joint density matrix according to \cite{breuer2002theory}
\begin{equation}
\begin{aligned}
    \dot{\rho} =
    &- i[H(t),\rho] + \Gamma_1{\cal D}[\sigma_-]\rho
    + \frac{\gamma_\phi}{2}{\cal D}[\sigma_z]\rho\\
    &+ \kappa(\bar n_{\rm th}+1){\cal D}[a_{\rm b}]\rho
    + \kappa\bar n_{\rm th}{\cal D}[a_{\rm b}^\dagger]\rho,
    \label{eq:master_equation}
\end{aligned}
\end{equation}
where ${\cal D}[L]\rho=L\rho L^\dagger -\{L^\dagger L,\rho\}/2$, $\Gamma_1=1/T_1$, and $1/T_2=\Gamma_1/2+\gamma_\phi$. The parameters \(T_1\) and \(T_2\) are the qubit energy-relaxation and transverse-coherence times, respectively. Energy relaxation gives \(\Gamma_1=1/T_1\), while \(T_2\) fixes the total decay rate of the qubit coherence. We separate this into relaxation-induced dephasing and pure
dephasing through \(\frac{1}{T_2}=\frac{\Gamma_1}{2}+\gamma_\phi\). Thus \((\gamma_\phi/2){\cal D}[\sigma_z]\rho\) accounts for dephasing not
caused by energy relaxation. The echo pulse cancels the deterministic
dynamical phase of the interferometer, but not this Markovian coherence loss. Unless stated otherwise, we use \(\Omega/2\pi=20\,{\rm MHz}\), \(\delta_0/2\pi=100\,{\rm MHz}\), \(\tau_p=25\,{\rm ns}\), and \(\tau_C=100\,{\rm ns}\). Qubit relaxation and dephasing are included with \(T_1=118\,{\rm ns}\) and \(T_2=157\,{\rm ns}\), corresponding to \(\Gamma_1=8.48\times10^{-3}\,{\rm ns}^{-1}\) and \(\gamma_\phi=2.13\times10^{-3}\,{\rm ns}^{-1}\). 
\begin{figure}
    \centering
    \includegraphics[width=\linewidth]{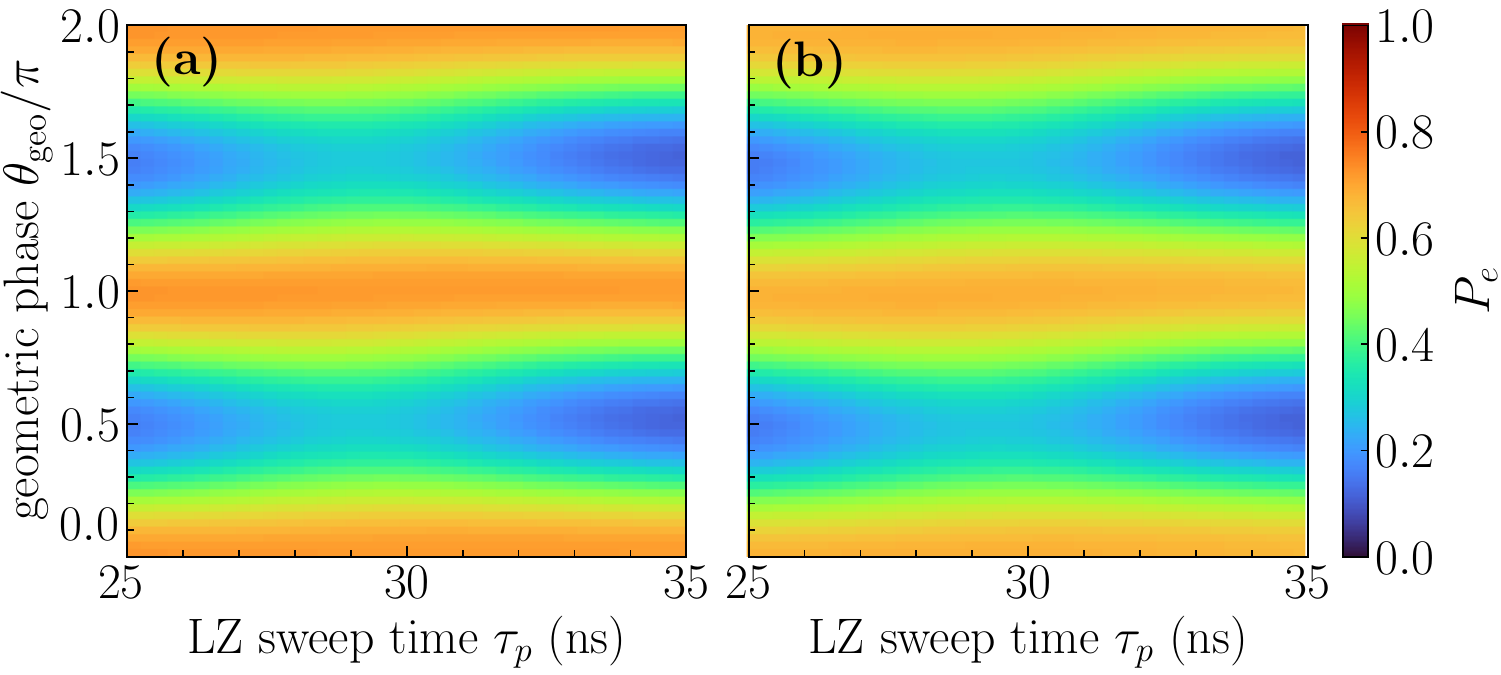}
    \caption{Finite-battery geometric Landau--Zener interference. The final excited-state population \(P_e=\operatorname{Tr}[\rho(\tau_C)|e\rangle\langle e|]\) is shown as a function of the geometric phase \(\theta_{\rm geo}\) and the Landau--Zener sweep time \(\tau_p\).
    (a) Classical-drive limit, obtained from the corresponding transverse-field Hamiltonian with the same mean gap
    \(\Omega/2\pi=20\,{\rm MHz}\).
    (b) Full quantum-battery evolution for a coherent battery with \(\bar n=5\), using \(g=\Omega/(2\sqrt{\bar n})\). Both panels use the same color scale.
    The finite battery preserves the echo-protected geometric fringe pattern, but the extrema are softened and weakly distorted by the photon-number-resolved gaps
    \(\Omega_n=\Omega\sqrt{n/\bar n}\), together with finite phase-reference strength and battery back-action.\justifying}
\label{fig:main_heatmaps}
\end{figure}

In the main figures we include a weak battery-loss channel with \(\kappa=10^{-4}\,{\rm ns}^{-1}\) and set \(\bar n_{\rm th}=0\). This corresponds to a resonator lifetime \(T_{\rm cav}=1/\kappa=10\,\mu{\rm s}\), so that the total loss accumulated during one interferometer cycle is small: for \(\tau_C=100\,{\rm ns}\), \(\kappa\tau_C=10^{-2}\). Equivalently, the mean photon number decays by only \(1-e^{-\kappa\tau_C}\simeq 1\%\), while the coherent amplitude, and hence the effective transverse gap, is reduced by only \(1-e^{-\kappa\tau_C/2}\simeq 0.5\%\). The parameters required for the quantum-battery implementation are otherwise modest on superconducting-circuit scales. For the representative coherent battery used in Fig.~\ref{fig:main_heatmaps}, \(\bar n=5\), the calibration \(g=\Omega/(2\sqrt{\bar n})\) gives \(g/2\pi\simeq 4.5\,{\rm MHz}\). Taking a typical microwave battery frequency \(\omega_b/2\pi\simeq 5\,{\rm GHz}\), this gives \(g/\omega_b\simeq 9\times10^{-4}\), well inside the rotating-wave regime. Thus the protocol duration is short compared with realistic resonator lifetimes, and the visible deviations from the classical geometric interferometer are not ordinary cavity-damping effects. They arise primarily from the finite quantum structure of the source: the coherent battery implements a superposition of photon-number-resolved Landau--Zener passages with gaps \(\Omega_n=\Omega\sqrt{n/\bar n}\).

During the two Jaynes--Cummings sweep segments, the lossless part of Eq.~\eqref{eq:H_QB} conserves the total excitation number \(N_{\rm tot}=a_{\rm b}^\dagger a_{\rm b}+|e\rangle\langle e|\). Hence the energy exchanged in the Landau--Zener passages is accounted for inside the qubit--battery Hilbert space, up to the explicitly included weak leakage channel \(\kappa\). The echo pulse is treated as an ideal qubit-only refocusing operation. It is not part of the battery-charging or battery-discharging step; rather, it removes the deterministic dynamical phase and turns the protocol into a geometric interferometer whose remaining fringe is controlled by the transverse phase supplied by the battery. With this separation, the finite-battery corrections reported below quantify the ability of a lossy but long-lived quantum battery to reproduce a phase-coherent classical control field.

We characterize the interferometer by the fringe contrast
$C=\max_{\theta_{\rm geo}}P_e(\theta_{\rm geo})
-\min_{\theta_{\rm geo}}P_e(\theta_{\rm geo})$, together with battery observables such as the photon-number change $\Delta n = \langle a_{\rm b}^\dagger a_{\rm b}\rangle_i-\langle a_{\rm b}^\dagger a_{\rm b}\rangle_f$, the photon-number variance ${\rm Var}(n)$, and the phase-reference fraction $\eta_{\rm coh}$. These quantities separate three physical effects of a finite quantum energy source: gap broadening, loss of phase-reference strength, and energetic back-action.

Figure~\ref{fig:main_heatmaps} shows the full two-parameter interferometric response. In the classical-drive limit, the echo-protected protocol produces the expected geometric fringe pattern: \(P_e\) oscillates with \(\theta_{\rm geo}\), while the dependence on \(\tau_p\) enters mainly through the Landau--Zener beam-splitter probability. Replacing the classical drive by a finite coherent battery with \(\bar n=5\) preserves the same phase-controlled structure, demonstrating that the quantized mode can power the geometric Landau--Zener interferometer. However, the finite-battery heatmap is not identical to the classical one. The extrema are slightly reduced and the pattern is weakly distorted because the coherent state contains a distribution of photon-number sectors, each realizing a different avoided-crossing gap \(\Omega_n=\Omega\sqrt{n/\bar n}\). Thus the measured signal is not produced by a single ideal Landau--Zener beam splitter, but by a coherent superposition of sector-dependent beam splitters. This provides the direct finite-battery signature in the interferometric output.

\begin{figure}
    \centering
    \includegraphics[width=\linewidth]{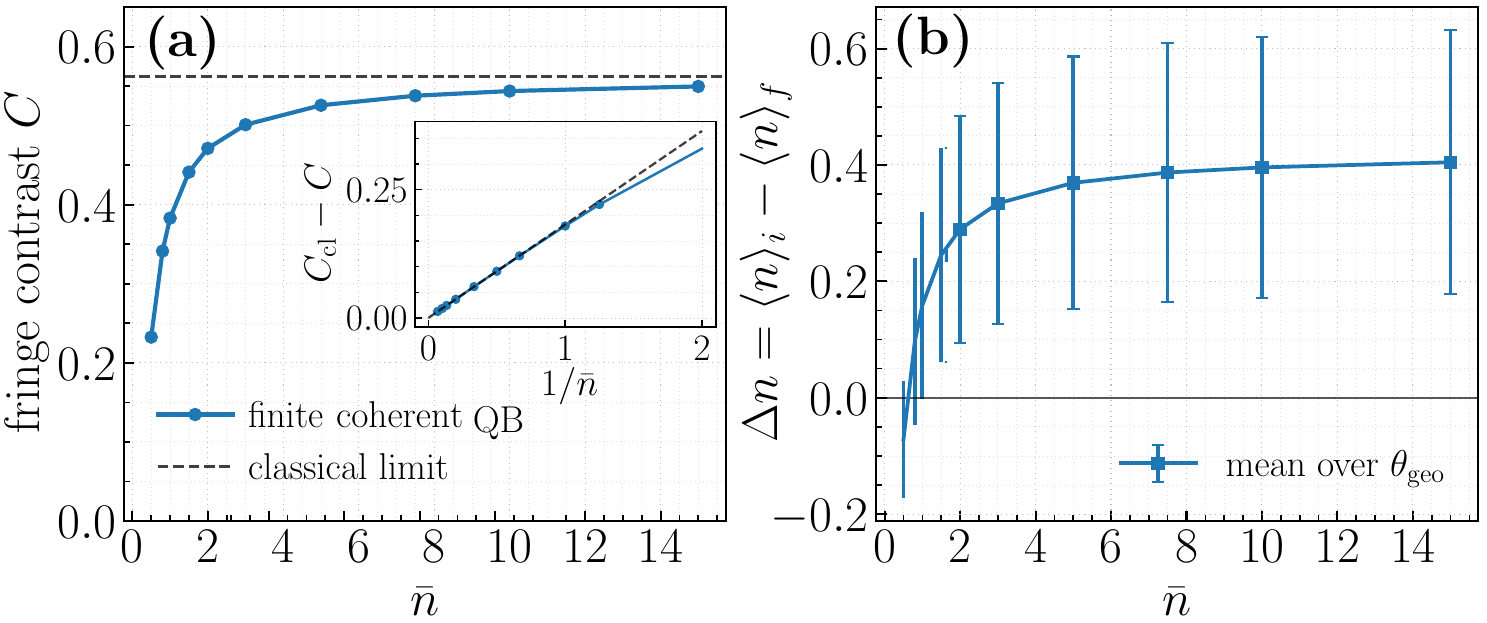}
    \caption{Quantum-to-classical recovery and battery back-action. (a) Fringe contrast \(C=\max_{\theta_{\rm geo}}P_e-\min_{\theta_{\rm geo}}P_e\) for a coherent quantum battery as a function of the mean photon number \(\bar n\). The dashed line gives the corresponding classical-drive limit with the same qubit relaxation and dephasing rates. The inset shows the contrast deficit \(C_{\rm cl}-C\) versus \(1/\bar n\), illustrating the approach to the classical limit as photon-number fluctuations become relatively small. (b) Battery back-action, quantified by the photon-number change \(\Delta n=\langle n\rangle_i-\langle n\rangle_f\), averaged over \(\theta_{\rm geo}\). Error bars denote the standard deviation over the geometric phase, not numerical uncertainty. The mean photon-number change remains of order one photon, while its relative size decreases with \(\bar n\). \justifying}
    \label{contrast_recovery}
\end{figure}

Figure~\figpanel{contrast_recovery}{a} quantifies how the finite coherent battery approaches the classical-drive limit. The fringe contrast increases monotonically with \(\bar n\) and tends toward the classical value, showing that the finite-battery distortions seen in Fig.~\ref{fig:main_heatmaps} are progressively suppressed as the coherent energy increases. This recovery is controlled by the relative photon-number fluctuations of the battery. Since a coherent state has \({\rm Var}(n)=\bar n\), the relative width of the sector-dependent gaps \(\Omega_n=\Omega\sqrt{n/\bar n}\) scales as \(1/\sqrt{\bar n}\), while the leading contrast deficit is well captured by a dependence on \(1/\bar n\), as shown in the inset (see Note~1 of the SM~\cite{SuppMat}).

This scaling follows from the Jaynes--Cummings matrix element. In a fixed photon-number sector, \(a_{\rm b}|n\rangle=\sqrt{n}|n-1\rangle\), so the states \(|n,g\rangle\) and \(|n-1,e\rangle\) experience an avoided-crossing gap \(\Omega_n=2g\sqrt{n}\). A battery with photon-number distribution \(p_n\) therefore generates a coherent superposition of sector-resolved Landau--Zener passages with different gaps. With \(g=\Omega/(2\sqrt{\bar n})\), one has \(\Omega_n=\Omega\sqrt{n/\bar n}\), and for \(n=\bar n+\delta n\) with \(|\delta n|\ll\bar n\),
\(
\frac{\delta\Omega_n}{\Omega}\simeq\frac{\delta n}{2\bar n}
\),
\(
\frac{\Delta\Omega}{\Omega}
\simeq
\frac{\sqrt{{\rm Var}(n)}}{2\bar n}.
\)
Thus photon-number fluctuations produce an inhomogeneous broadening of the Landau--Zener gap. For a coherent battery, \({\rm Var}(n)=\bar n\), so the relative gap width decreases as \(1/(2\sqrt{\bar n})\), while the leading contrast deficit scales as \(1/\bar n\) in the adiabatic-impulse estimate (see Note~1 of the SM~\cite{SuppMat}).

The battery is nevertheless not a passive classical parameter. Panel~\figpanel{contrast_recovery}{b} shows the photon-number change after one interferometric cycle, averaged over the geometric phase. The mean photon-number change remains finite, reflecting genuine energy exchange and back-action between the qubit and the QB. However, this change is small compared with the total stored energy when \(\bar n\) is large, so the relative back-action \(\Delta n/\bar n\) vanishes in the higher coherent energy limit. Thus the same calculation captures both limits: for small \(\bar n\), the battery is a finite quantum resource whose number fluctuations and back-action affect the interferometer; for large \(\bar n\), it behaves as an effectively classical phase-coherent drive.

The finite-battery problem, however, is not only a question of number fluctuations. Geometric Landau--Zener interference also requires a phase reference, because the geometric phase is controlled by the azimuthal direction of the transverse field. In the present model this reference is supplied by the coherent amplitude of the battery. We quantify it by \(\eta_{\rm coh} = |\langle a_{\rm b}\rangle|^2/\langle a_{\rm b}^\dagger a_{\rm b}\rangle=|\langle a_{\rm b}\rangle|^2/\bar n\), which gives the classical-like transverse gap \(\Omega_{\rm eff} = 2g|\langle a_{\rm b}\rangle| = \Omega\sqrt{\eta_{\rm coh}}\). The competition is especially transparent for a weak amplitude-squeezed coherent battery. At fixed total energy \(\bar n\), Note~2 of the SM gives
\(
{\rm Var}_{\rm amp}(n)
\simeq
\bar n-2\bar n r+(2\bar n+1)r^2
\),
\(
\eta_{\rm coh}
\simeq
1-\frac{r^2}{\bar n},
\)
up to higher-order corrections in \(r\) (see Note 2 of the SM~\cite{SuppMat}). Thus weak amplitude squeezing narrows the photon-number distribution already at first order in \(r\), while the loss of coherent phase-reference strength starts only at second order. This suggests a possible sweet spot: reducing \({\rm Var}(n)\) suppresses the spread of sector-dependent gaps, whereas reducing \(\eta_{\rm coh}\) weakens the phase reference that controls the geometric interference.

The full quantum-battery dynamics shows, however, that this favorable small-\(r\) counting is not by itself sufficient. For the present GLZI protocol and at fixed total battery energy, the coherent state gives the largest fringe contrast among the tested states (see Note~2 of the SM~\cite{SuppMat}). Although amplitude-squeezed and number-squeezed states reduce photon-number fluctuations and hence narrow the distribution of sector-dependent gaps \(\Omega_n\), this benefit is outweighed in the scanned regime by the reduction of the coherent phase-reference fraction. Thus the contrast is controlled not only by the sharpness of the photon-number distribution, but also, and more strongly here, by the availability of first-order phase coherence in the battery. The relevant resource is therefore not stored energy alone, but phase-coherent battery energy. Phase squeezing is even less favorable for this purpose, since it increases photon-number fluctuations while still reducing the coherent fraction. This distinguishes geometric control from energy-exchange gates powered by number states~\cite{l39v-jwwz}: a Fock-state battery can supply a sharply defined quantum of energy, but because \(\langle a_{\rm b}\rangle=0\) it does not by itself provide the first-order phase reference required to set the geometric-control axis.

An important practical implication is that the interferometric signal does not require a macroscopically populated control field. Even for a coherent battery with mean occupation of only a few quanta, the geometric fringe remains clearly visible. For example, already at \(\bar n=2\) the contrast is a sizable fraction of the classical-drive value, and the full \((\theta_{\rm geo},\tau_p)\) heatmap retains the echo-protected geometric-fringe structure (see Fig.~2 of the SM~\cite{SuppMat}). This few-quanta regime is conceptually distinct from the usual classical-drive description, where the microwave field is treated as an effectively infinite phase-coherent reservoir. In cryogenic hardware, externally supplied classical microwave fields are accompanied by input-line attenuation, filtering, wiring overhead, and heat load. Our result shows that the phase information required for geometric Landau--Zener control can, in principle, be supplied by phase-coherent energy stored in a finite quantum mode. The relevant resource is therefore not a large photon number, but a sufficiently stable coherent amplitude \( \langle a_{\rm b}\rangle \).

\emph{Conclusions}---
Classical microwave fields are usually treated as ideal phase-coherent work sources for qubit control. Here we asked what changes when such a field is replaced by a finite quantum battery and tested this question in a geometric Landau--Zener interferometer. The quantized battery does not implement a single classical beam splitter. Instead, the Jaynes--Cummings matrix element decomposes each sweep segment into
photon-number sectors, while the qubit-only echo couples neighboring sector labels
between the two passages. A coherent battery therefore recovers the classical-drive result only in the macroscopic limit, while finite \(\bar n\) produces softened fringes, photon-number-resolved distortions, and measurable back-action on the battery.
The central lesson is not simply that finite batteries reduce contrast. Rather, geometric Landau--Zener interferometry certifies whether an internal quantum energy source supplies phase-coherent energy. Stored energy alone is insufficient: the same mode must provide the transverse phase reference that fixes the geometric-control axis. This is why reducing photon-number fluctuations is not automatically beneficial; it can come at the cost of reduced \(\eta_{\rm coh}=|\langle a_{\rm b}\rangle|^2/\bar n\). The protocol therefore separates three resources that are hidden inside an ideal classical drive: energy, phase reference, and back-action. In this sense, geometric Landau--Zener interferometry provides a practical benchmark for quantum-battery-powered control, certifying not only that energy is available, but that it is available in a phase-coherent form useful for quantum operations. The persistence of visible fringes at few-quanta occupation suggests a route toward low-power, internally powered coherent control, where the control resource is phase-coherent stored energy rather than a macroscopic externally supplied field.

\acknowledgments{\emph{Acknowledgments}---BA thanks Pertti Hakonen and Konrad Schlichtholz for fruitful discussions. BA acknowledges support from IRA Program (project no. FENG.02.01-IP.05-0006/23) financed by the FENG program 2021-2027, Priority FENG.02, Measure FENG.02.01., with the support of the FNP.}

\clearpage
\onecolumngrid  


\setcounter{section}{0}
\setcounter{subsection}{0}
\setcounter{equation}{0}
\setcounter{figure}{0}
\setcounter{table}{0}

\renewcommand{\theequation}{S\arabic{equation}}
\renewcommand{\thefigure}{\arabic{figure}}
\renewcommand{\thetable}{S\arabic{table}}

\newcommand{\suppnote}[1]{%
    \refstepcounter{section}%
    \setcounter{subsection}{0}%
    \section*{Supplementary Note \arabic{section}. #1}%
    \addcontentsline{toc}{section}{Supplementary Note \arabic{section}. #1}%
}

\newcommand{\suppsubsection}[1]{%
    \refstepcounter{subsection}%
    \subsection*{\Alph{subsection}. #1}%
    \addcontentsline{toc}{subsection}{\Alph{subsection}. #1}%
}

\suppnote{Photon-number-induced broadening of the Landau--Zener gap}
\label{sec:SM_gap_broadening}

In this section we derive the finite-battery broadening mechanism used in the main text and clarify its domain of validity in the presence of the qubit-only echo pulse. The key point is that the Jaynes--Cummings Hamiltonian decomposes into photon-number-resolved two-level blocks during each Landau--Zener sweep segment. The echo pulse, however, is not generated by the battery Hamiltonian and does not preserve the corresponding excitation-sector label. Thus \(\Omega_n=2g\sqrt n\) is an exact dressed-gap formula for each sweep segment, while the complete echo-refocused interferometer must be evolved in the full qubit--battery Hilbert space.

During each Landau--Zener sweep segment, the Hamiltonian is
\begin{equation}
    H(t)
    =
    \frac{\delta(t)}{2}\sigma_z
    +
    g\left(
        a_{\rm b}\sigma_+
        +
        a_{\rm b}^{\dagger}\sigma_-
    \right),
    \label{eq:SM_JC_H}
\end{equation}
where \(a_{\rm b}\) annihilates a battery excitation and
\(\sigma_+=|e\rangle\langle g|\), \(\sigma_-=|g\rangle\langle e|\).
Equation~\eqref{eq:SM_JC_H} commutes with the total-excitation operator
\begin{equation}
    N_{\rm tot}
    =
    a_{\rm b}^{\dagger}a_{\rm b}
    +
    |e\rangle\langle e| .
    \label{eq:SM_Ntot}
\end{equation}
Therefore, between echo operations, the Hilbert space decomposes into independent Jaynes--Cummings blocks. For each integer \(n\ge 1\), the subspace spanned by
\(\{|n,g\rangle,|n-1,e\rangle\}\) is invariant under Eq.~\eqref{eq:SM_JC_H}. The relevant matrix element is
\begin{equation}
    \langle n-1,e|
    g a_{\rm b}\sigma_+
    |n,g\rangle
    =
    g\sqrt n ,
    \label{eq:SM_JC_matrix_element}
\end{equation}
because \(a_{\rm b}|n\rangle=\sqrt n\,|n-1\rangle\). Thus, in the basis
\(\{|n-1,e\rangle,|n,g\rangle\}\), the Hamiltonian in the \(n\)th excitation block is
\begin{equation}
    H_n(t)
    =
    \begin{pmatrix}
    \delta(t)/2 & g\sqrt n \\
    g\sqrt n & -\delta(t)/2
    \end{pmatrix}
    =
    \frac{\delta(t)}{2}\tau_z
    +
    g\sqrt n\,\tau_x ,
    \label{eq:SM_Hn_matrix}
\end{equation}
where \(\tau_x\) and \(\tau_z\) are Pauli operators acting inside this two-dimensional block. Comparing Eq.~\eqref{eq:SM_Hn_matrix} with the standard Landau--Zener form
\begin{equation}
    H_{\rm LZ}(t)
    =
    \frac{\delta(t)}{2}\tau_z
    +
    \frac{\Omega}{2}\tau_x ,
    \label{eq:SM_H_LZ_standard}
\end{equation}
identifies the photon-number-dependent avoided-crossing gap
\begin{equation}
    \Omega_n
    =
    2g\sqrt n .
    \label{eq:SM_Omega_n}
\end{equation}
This is the microscopic origin of the first finite-battery correction: different photon-number components of the battery generate different Landau--Zener gaps during a sweep.

The block decomposition above is exact only between instantaneous echo operations. The echo pulse used in the main text is a qubit-only operation,
\begin{equation}
    U_\pi(\phi_{\rm echo})
    =
    \exp\left[
        -\frac{i\pi}{2}
        \left(
            \cos\phi_{\rm echo}\,\sigma_x
            +
            \sin\phi_{\rm echo}\,\sigma_y
        \right)
    \right],
    \label{eq:SM_echo_unitary}
\end{equation}
acting on the full Hilbert space as \(\mathbb{I}_B\otimes U_\pi\). Using
\begin{equation}
    \cos\phi\,\sigma_x+\sin\phi\,\sigma_y
    =
    e^{-i\phi}\sigma_+
    +
    e^{i\phi}\sigma_- ,
    \label{eq:SM_equatorial_axis}
\end{equation}
one obtains
\begin{equation}
    U_\pi(\phi_{\rm echo})
    =
    -i
    \left(
        e^{-i\phi_{\rm echo}}\sigma_+
        +
        e^{i\phi_{\rm echo}}\sigma_-
    \right).
    \label{eq:SM_echo_sigma_pm}
\end{equation}
Therefore
\begin{equation}
    U_\pi(\phi_{\rm echo})|g\rangle
    =
    -i e^{-i\phi_{\rm echo}}|e\rangle,
    \qquad
    U_\pi(\phi_{\rm echo})|e\rangle
    =
    -i e^{i\phi_{\rm echo}}|g\rangle .
    \label{eq:SM_echo_action_qubit}
\end{equation}
On the joint qubit--battery basis states this gives
\begin{equation}
    |n,g\rangle
    \xrightarrow{\ \mathbb{I}_B\otimes U_\pi\ }
    -i e^{-i\phi_{\rm echo}} |n,e\rangle ,
    \label{eq:SM_echo_ng}
\end{equation}
and
\begin{equation}
    |n-1,e\rangle
    \xrightarrow{\ \mathbb{I}_B\otimes U_\pi\ }
    -i e^{i\phi_{\rm echo}} |n-1,g\rangle .
    \label{eq:SM_echo_nminus1e}
\end{equation}
These final states no longer belong to the original \(n\)-excitation Jaynes--Cummings block. Indeed,
\begin{equation}
    N_{\rm tot}|n,g\rangle=n|n,g\rangle,
    \qquad
    N_{\rm tot}|n-1,e\rangle=n|n-1,e\rangle,
    \label{eq:SM_N_before_echo}
\end{equation}
whereas
\begin{equation}
    N_{\rm tot}|n,e\rangle=(n+1)|n,e\rangle,
    \qquad
    N_{\rm tot}|n-1,g\rangle=(n-1)|n-1,g\rangle .
    \label{eq:SM_N_after_echo}
\end{equation}
Equivalently,
\begin{equation}
    \left[
        \mathbb{I}_B\otimes U_\pi(\phi_{\rm echo}),
        N_{\rm tot}
    \right]
    \neq 0 .
    \label{eq:SM_echo_not_conserve_N}
\end{equation}
Thus the echo maps amplitudes from the \(n\)-sector into neighboring sectors before the second sweep.

Consequently, an amplitude evolving in the \(n\)-sector during the first Landau--Zener passage generally continues after the echo in neighboring Jaynes--Cummings sectors. The relevant gaps are then
\begin{equation}
    \Omega_{n+1}=2g\sqrt{n+1},
    \qquad
    \Omega_{n-1}=2g\sqrt{n-1},
    \label{eq:SM_neighboring_gaps}
\end{equation}
rather than the original \(\Omega_n=2g\sqrt n\). For large \(n\), this sector shift is asymptotically small. Indeed,
\begin{equation}
    \Omega_{n\pm1}
    =
    2g\sqrt{n\pm1}
    =
    \Omega_n
    \sqrt{1\pm\frac{1}{n}} .
    \label{eq:SM_neighbor_gap_start}
\end{equation}
Using
\begin{equation}
    \sqrt{1+x}
    =
    1+\frac{x}{2}-\frac{x^2}{8}+O(x^3),
    \qquad |x|\ll1 ,
    \label{eq:SM_sqrt_expansion_echo}
\end{equation}
we find
\begin{equation}
    \Omega_{n+1}
    =
    \Omega_n
    \left(
        1+\frac{1}{2n}
        -
        \frac{1}{8n^2}
        +
        O(n^{-3})
    \right),
    \label{eq:SM_Omega_np1_expanded}
\end{equation}
and
\begin{equation}
    \Omega_{n-1}
    =
    \Omega_n
    \left(
        1-\frac{1}{2n}
        -
        \frac{1}{8n^2}
        +
        O(n^{-3})
    \right).
    \label{eq:SM_Omega_nm1_expanded}
\end{equation}
In compact form,
\begin{equation}
    \Omega_{n\pm1}
    =
    \Omega_n
    \left(
        1
        \pm
        \frac{1}{2n}
        +
        O(n^{-2})
    \right).
    \label{eq:SM_neighbor_gap_compact}
\end{equation}
The echo-induced sector redistribution therefore gives relative corrections of order \(1/n\). For large \(n\), the same-sector picture becomes asymptotically accurate, but for few-quanta batteries this sector shift is part of the finite-battery physics. Accordingly, Eq.~\eqref{eq:SM_Omega_n} should be understood as an exact statement about each Jaynes--Cummings passage between echo operations, not as an exact independent-sector decomposition of the complete echo-refocused interferometer. All quantitative results in the main text are obtained from the full qubit--battery density-matrix evolution, which includes this echo-induced redistribution exactly.

We now use the sector-dependent gap formula to estimate the broadening caused by the photon-number distribution of the battery. For a coherent battery with mean photon number \(\bar n\), the coupling is calibrated as
\(g=\Omega/(2\sqrt{\bar n})\), so that the central photon-number sector has the nominal gap \(\Omega\). With this convention, Eq.~\eqref{eq:SM_Omega_n} becomes
\begin{equation}
    \Omega_n
    =
    \Omega
    \sqrt{\frac{n}{\bar n}} .
    \label{eq:SM_Omega_n_scaled}
\end{equation}
Writing \(n=\bar n+\delta n\), we obtain
\begin{equation}
    \Omega_n
    =
    \Omega
    \sqrt{
        1+\frac{\delta n}{\bar n}
    } .
    \label{eq:SM_Omega_expansion_start}
\end{equation}
For photon-number fluctuations small compared with the mean,
\(|\delta n|\ll \bar n\), expanding the square root gives
\begin{equation}
    \Omega_n
    =
    \Omega
    \left[
        1
        +
        \frac{\delta n}{2\bar n}
        -
        \frac{\delta n^2}{8\bar n^2}
        +
        O\left(
            \frac{\delta n^3}{\bar n^3}
        \right)
    \right].
    \label{eq:SM_Omega_n_expanded}
\end{equation}
To leading order,
\begin{equation}
    \frac{\delta\Omega_n}{\Omega}
    =
    \frac{\Omega_n-\Omega}{\Omega}
    \simeq
    \frac{\delta n}{2\bar n}.
    \label{eq:SM_relative_gap_shift}
\end{equation}
Therefore, if the photon-number distribution has variance
\begin{equation}
    {\rm Var}(n)
    =
    \langle n^2\rangle
    -
    \langle n\rangle^2 ,
    \label{eq:SM_var_n}
\end{equation}
the root-mean-square width of the gap distribution is
\begin{equation}
    \frac{\Delta\Omega}{\Omega}
    \simeq
    \frac{\sqrt{{\rm Var}(n)}}{2\bar n}.
    \label{eq:SM_relative_gap_width}
\end{equation}
Equation~\eqref{eq:SM_relative_gap_width} is the broadening formula quoted in the main text. For a coherent state,
\begin{equation}
    p_n
    =
    e^{-\bar n}
    \frac{\bar n^n}{n!},
    \qquad
    {\rm Var}(n)=\bar n ,
    \label{eq:SM_coherent_distribution}
\end{equation}
so that
\begin{equation}
    \frac{\Delta\Omega}{\Omega}
    \simeq
    \frac{1}{2\sqrt{\bar n}} .
    \label{eq:SM_coherent_gap_width}
\end{equation}
Even a coherent battery therefore produces a finite spread of Landau--Zener gaps. This spread vanishes only in the classical-drive limit \(\bar n\rightarrow\infty\).

The same mechanism can be expressed in terms of the Landau--Zener probability. With the convention that the minimum avoided-crossing gap is \(\Omega=2\Delta\), the standard asymptotic Landau--Zener probability used in superconducting LZ interferometry is
\cite{Zener1932,Shevchenko2010,PhysRevLett.96.187002}
\begin{equation}
    P_{\rm LZ}^{(n)}
    =
    \exp\left[
        -\frac{\pi\Omega_n^2}{2v}
    \right].
    \label{Landau-Zener_PLZ}
\end{equation}
This expression is used only to expose the scaling with photon-number fluctuations. In the finite-time GLZI protocol, all quantitative results are obtained from the full qubit--battery dynamics. Using Eq.~\eqref{eq:SM_Omega_n_scaled}, Eq.~\eqref{Landau-Zener_PLZ} becomes
\begin{equation}
    P_{\rm LZ}^{(n)}
    =
    \exp
    \left[
        -\frac{\pi\Omega^2}{2v}
        \frac{n}{\bar n}
    \right].
    \label{eq:SM_PLZ_n_scaled}
\end{equation}
If
\begin{equation}
    P_{\rm LZ}^{(0)}
    =
    \exp
    \left[
        -\frac{\pi\Omega^2}{2v}
    \right]
    \label{eq:SM_PLZ_classical}
\end{equation}
denotes the Landau--Zener probability associated with the mean-field gap \(\Omega\), then
\begin{equation}
    P_{\rm LZ}^{(n)}
    =
    \left(
        P_{\rm LZ}^{(0)}
    \right)^{n/\bar n}.
    \label{eq:SM_PLZ_power_law}
\end{equation}
Thus a battery state with photon-number distribution \(p_n\) does not realize a single Landau--Zener passage with one fixed gap. Instead, the qubit signal contains contributions from sector-resolved passages with different Landau--Zener probabilities. In the full echo-refocused protocol these sectors are also redistributed by the qubit-only echo, as described above.

In the ideal adiabatic-impulse estimate, a sector with gap \(\Omega_n\) would give a geometric fringe \cite{PhysRevLett.112.027001,PhysRevA.73.063405,science.1119678}
\begin{equation}
    P_e^{(n)}(\theta_{\rm geo})
    =
    1
    -
    A_n
    \sin^2\theta_{\rm geo},
    \qquad
    A_n
    =
    4P_{\rm LZ}^{(n)}
    \left(
        1-P_{\rm LZ}^{(n)}
    \right).
    \label{eq:SM_sector_fringe}
\end{equation}
This estimate is not used as the numerical model; it only gives the leading scaling of the contrast deficit. For a narrow distribution centered at \(\bar n\), let \(x=n/\bar n\) and define
\(\beta=\pi\Omega^2/(2v)\), so that \(P_{\rm LZ}^{(n)}=e^{-\beta x}\). Then
\begin{equation}
    A(x)
    =
    4e^{-\beta x}
    \left(
        1-e^{-\beta x}
    \right).
    \label{eq:SM_Ax}
\end{equation}
Since \(\langle x-1\rangle=0\), the leading correction to the averaged amplitude is second order in the photon-number width:
\begin{equation}
    \langle A\rangle
    \simeq
    A(1)
    +
    \frac{1}{2}
    A''(1)
    \frac{{\rm Var}(n)}{\bar n^2}.
    \label{eq:SM_A_average_expansion}
\end{equation}
A direct differentiation gives
\begin{equation}
    A''(1)
    =
    4\beta^2
    P_{\rm LZ}^{(0)}
    \left(
        1
        -
        4P_{\rm LZ}^{(0)}
    \right).
    \label{eq:SM_A_second_derivative}
\end{equation}
Therefore, in the high-contrast operating regime \(P_{\rm LZ}^{(0)}>1/4\), photon-number broadening lowers the averaged fringe amplitude:
\begin{equation}
    A(1)-\langle A\rangle
    \simeq
    2\beta^2
    P_{\rm LZ}^{(0)}
    \left(
        4P_{\rm LZ}^{(0)}-1
    \right)
    \frac{{\rm Var}(n)}{\bar n^2}.
    \label{eq:SM_contrast_deficit_general}
\end{equation}
For a coherent battery, \({\rm Var}(n)=\bar n\), so the leading contrast deficit scales as
\begin{equation}
    A(1)-\langle A\rangle
    \propto
    \frac{1}{\bar n}.
    \label{eq:SM_contrast_deficit_coherent_scaling}
\end{equation}
This explains the asymptotic recovery of the classical fringe with a leading \(1/\bar n\) correction. The derivation above isolates only the gap-broadening mechanism. It does not assume that photon-number noise is the only finite-battery effect. In the full quantum evolution, the battery can become entangled with the qubit, the echo can redistribute amplitudes between excitation sectors, and nonclassical battery states can also change the phase-reference strength \(|\langle a_{\rm b}\rangle|^2/\bar n\). The purpose of Eq.~\eqref{eq:SM_relative_gap_width} is therefore diagnostic: it shows why reducing \({\rm Var}(n)\) can suppress the spread of Landau--Zener gaps, while the main text explains why this improvement must be balanced against the possible loss of coherent phase reference.
\begin{figure*}
    \centering
    \includegraphics[width=0.7\textwidth]{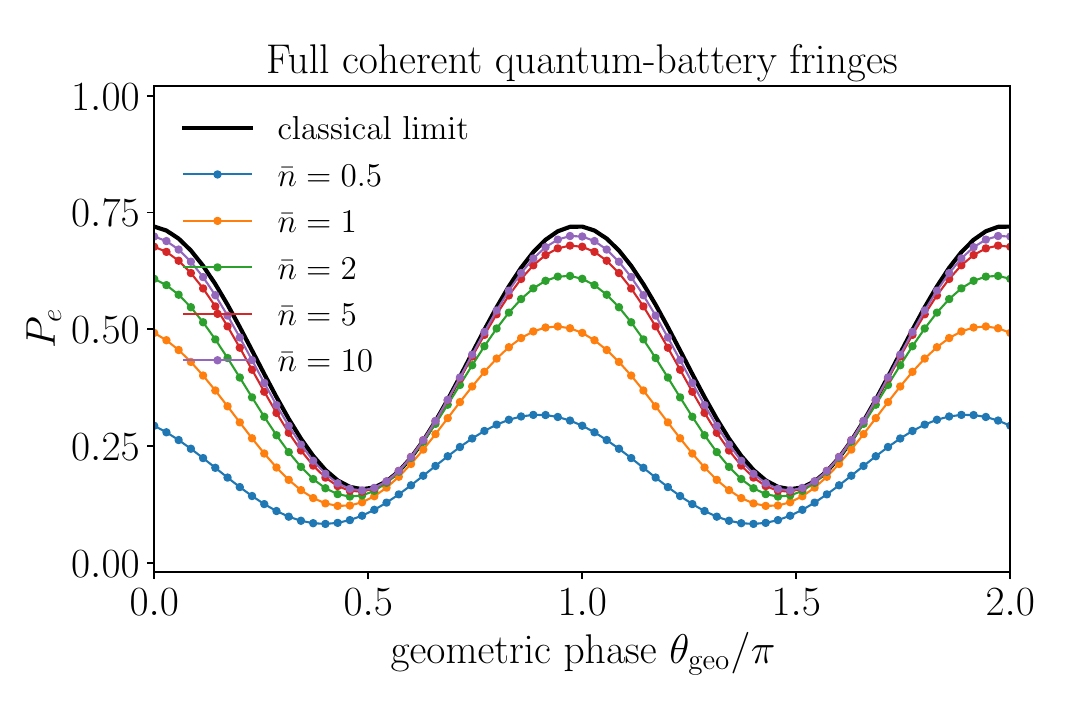}
    \caption{
    Coherent quantum-battery fringes for different mean photon numbers.
    The final excited-state population \(P_e\) is shown as a function of the geometric phase
    \(\theta_{\rm geo}\) for coherent batteries with \(\bar n=0.5,1,2,5,10\), together with the classical-drive limit.
    All curves are obtained from the full qubit--battery master equation, with the coupling calibrated as
    \(g=\Omega/(2\sqrt{\bar n})\) so that the central photon-number sector has the same nominal gap \(\Omega\).
    As \(\bar n\) increases, the finite-battery fringe approaches the classical response.
    At small \(\bar n\), the fringe contrast is reduced because the coherent state contains a broad distribution of photon-number components, each generating a different Landau--Zener gap during a sweep, and because the qubit dynamics produces appreciable battery back-action.
    The recovery of the classical curve at large \(\bar n\) reflects the emergence of a macroscopic phase-coherent drive.
    \justifying}
    \label{fig6_quantum_fringe_grid}
\end{figure*}

Figure~\ref{fig6_quantum_fringe_grid} shows phase-fringe line cuts for coherent quantum batteries with increasing mean photon number. The classical-drive limit is recovered progressively as \(\bar n\) grows. This behavior follows from the photon-number-resolved structure of the Jaynes--Cummings coupling during each sweep segment. A coherent state with mean photon number \(\bar n\) contains a distribution of Fock components, and each component generates a Landau--Zener passage with a sector-dependent gap \(\Omega_n\).

For small \(\bar n\), the relative width of the photon-number distribution is large, so the observed signal contains contributions from significantly different Landau--Zener beam splitters. This produces a softened fringe and a reduced contrast. As \(\bar n\) increases, the relative number fluctuations decrease, the distribution of \(\Omega_n\) becomes narrower, and the finite-battery fringe converges toward the classical response. The remaining difference at finite \(\bar n\) is therefore a direct signature of the quantized nature of the battery mode rather than a numerical artifact or a mean-field effect. This line-cut representation complements the heatmaps in the main text and isolates the same quantum-to-classical recovery at fixed sweep time.
\begin{figure}[t]
    \centering
    \includegraphics[width=0.7\linewidth]{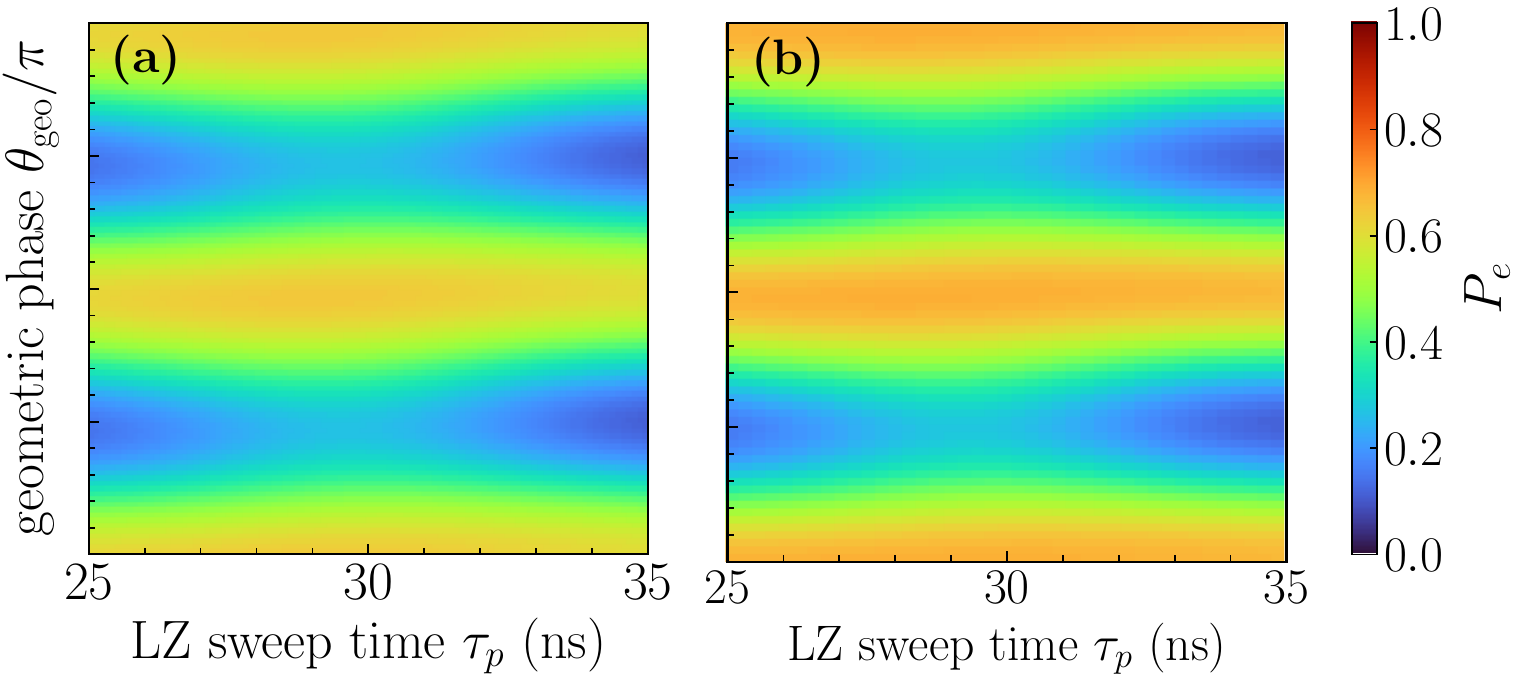}
    \caption{Finite-coherent-battery heatmaps at few-quanta occupation.
    The final excited-state population \(P_e\) is shown as a function of the geometric phase \(\theta_{\rm geo}\) and the Landau--Zener sweep time \(\tau_p\), using the full qubit--battery master equation.
    Panels show coherent batteries with (a) \(\bar n=2\) and (b) \(\bar n=5\), with the coupling calibrated as \(g=\Omega/(2\sqrt{\bar n})\) in each case so that the central photon-number sector has the same nominal gap \(\Omega\).
    All other parameters are the same as in the main text: \(\Omega/2\pi=20\,{\rm MHz}\), \(\delta_0/2\pi=100\,{\rm MHz}\), \(\tau_C=100\,{\rm ns}\), \(T_1=118\,{\rm ns}\), \(T_2=157\,{\rm ns}\), \(\kappa=10^{-4}\,{\rm ns}^{-1}\), and \(\bar n_{\rm th}=0\).
    The geometric fringe pattern is already visible for \(\bar n=2\), demonstrating that a macroscopically occupied drive is not required.
    Increasing the coherent battery energy to \(\bar n=5\) reduces the relative photon-number width, narrows the distribution of sector-dependent gaps \(\Omega_n=\Omega\sqrt{n/\bar n}\), and gives a heatmap closer to the classical-drive response.\justifying}
    \label{fig:SM_finite_QB_heatmaps_n2_n5}
\end{figure}

Figure~\ref{fig:SM_finite_QB_heatmaps_n2_n5} gives the two-dimensional version of the few-quanta recovery discussed in the main text. Both panels are obtained from the full finite-QB dynamics, not from a mean-field replacement of the battery operator. The persistence of the horizontal geometric-fringe bands at \(\bar n=2\) shows that the phase information required for echo-refocused geometric Landau--Zener interferometry can already be supplied by a small coherent battery. The pattern is nevertheless softer than for larger \(\bar n\), because the coherent state has a broad photon-number distribution relative to its mean. Each photon-number component generates a different avoided-crossing gap during the Jaynes--Cummings sweep, so the measured signal reflects sector-resolved finite-battery dynamics rather than a single classical beam splitter.

Increasing the mean occupation from \(\bar n=2\) to \(\bar n=5\) reduces the relative photon-number fluctuations and therefore the inhomogeneous broadening of the Landau--Zener beam splitters. This sharpens the extrema of \(P_e\) while preserving the same phase-controlled structure. The comparison confirms that the quantum-to-classical recovery is not only visible in one-dimensional line cuts, but also in the full \((\theta_{\rm geo},\tau_p)\) interferogram. It also supports the main-text interpretation: the dominant finite-battery correction is the photon-number-resolved gap structure of the Jaynes--Cummings coupling, while the weak cavity loss used here, \(\kappa\tau_C=10^{-2}\), produces only percent-level damping over one interferometer cycle.

\suppnote{Initial squeezed battery states}
\label{sec:sm_squeezed_battery}

In this section we analyze the role of initially squeezed battery states in the quantum-battery implementation of geometric Landau--Zener interferometry. The purpose is to separate two effects that are both important for a finite bosonic energy source. First, the photon-number distribution determines the spread of Landau--Zener gaps,
\begin{equation}
    \Omega_n = 2g\sqrt{n}.
\end{equation}
Second, the phase coherence of the battery determines how well the battery can act as a phase reference for the transverse qubit coupling. Squeezing can reduce number fluctuations, but at fixed total energy it can also reduce the coherent displacement that supplies this phase reference.

We consider the closed qubit--battery Hamiltonian
\begin{equation}
    H(t)
    =
    \frac{\delta(t)}{2}\sigma_z
    +
    g\left(a_{\rm b}\sigma_+ + a_{\rm b}^\dagger\sigma_-\right),
    \label{eq:sm_jc_hamiltonian}
\end{equation}
where \(a_{\rm b}\) annihilates a battery excitation and \(\sigma_+=\lvert e\rangle\langle g\rvert\), \(\sigma_-=\lvert g\rangle\langle e\rvert\). The total excitation number
\begin{equation}
    N_{\rm tot}
    =
    a_{\rm b}^\dagger a_{\rm b}
    +
    \lvert e\rangle\langle e\rvert
\end{equation}
is conserved by Eq.~\eqref{eq:sm_jc_hamiltonian}. Thus the Hamiltonian decomposes into independent two-dimensional sectors (between echo operations) spanned by
\begin{equation}
    \left\{
    \lvert n,g\rangle,\,
    \lvert n-1,e\rangle
    \right\},
    \qquad n\geq 1.
\end{equation}
In this sector,
\begin{equation}
    H_n(t)
    =
    \begin{pmatrix}
    -\delta(t)/2 & g\sqrt{n} \\
    g\sqrt{n} & \delta(t)/2
    \end{pmatrix},
    \label{eq:sm_sector_hamiltonian}
\end{equation}
so the instantaneous dressed energies are
\begin{equation}
    E_\pm^{(n)}(t)
    =
    \pm
    \frac{1}{2}
    \sqrt{\delta(t)^2+\Omega_n^2},
    \qquad
    \Omega_n=2g\sqrt{n}.
    \label{eq:sm_sector_energies}
\end{equation}
If the coupling is calibrated according to
\begin{equation}
    g=\frac{\Omega}{2\sqrt{\bar n}},
\end{equation}
then
\begin{equation}
    \Omega_n
    =
    \Omega\sqrt{\frac{n}{\bar n}}.
\label{eq:sm_gap_spread}
\end{equation}
Therefore a battery state with photon-number distribution \(p_n\) realizes a coherent superposition of photon-number-resolved Landau--Zener passages with different gaps. For a distribution narrow around \(\bar n\),
\begin{equation}
    \frac{\delta \Omega_n}{\Omega}
    =
    \sqrt{\frac{n}{\bar n}}-1
    \simeq
    \frac{n-\bar n}{2\bar n},
\end{equation}
and hence
\begin{equation}
    \frac{\operatorname{Var}(\Omega_n)}{\Omega^2}
    \simeq
    \frac{\operatorname{Var}(n)}{4\bar n^2}.
    \label{eq:sm_gap_variance}
\end{equation}
This relation is the main reason why number squeezing can, in principle, improve the interference contrast: it narrows the distribution of Landau--Zener beam splitters.

\suppsubsection{Closed unitary evolution for a general initial battery state}

Let the battery initially be in an arbitrary pure state
\begin{equation}
    \lvert \psi_B\rangle
    = \sum_{n=0}^{\infty} c_n\lvert n\rangle,
    \label{eq:sm_general_battery_state}
\end{equation}
and let the qubit initially be in \(\lvert g\rangle\). Then
\begin{equation}
    \lvert \Psi(0)\rangle
    =
    \sum_{n=0}^{\infty}c_n\lvert n,g\rangle.
\end{equation}
Because the Hamiltonian conserves \(N_{\rm tot}\), the exact state can be written as
\begin{equation}
    \lvert \Psi(t)\rangle
    =
    c_0 A_0(t)\lvert 0,g\rangle
    +
    \sum_{n=1}^{\infty}
    c_n
    \left[
    A_n(t)\lvert n,g\rangle
    +
    B_n(t)\lvert n-1,e\rangle
    \right],
    \label{eq:sm_exact_sector_evolution}
\end{equation}
where $A_0(t)=1$ and
\begin{equation}
    \begin{pmatrix}
    A_n(t)\\
    B_n(t)
    \end{pmatrix}
    =
    U_n(t)
    \begin{pmatrix}
    1\\
    0
    \end{pmatrix},
\end{equation}
and
\begin{equation}
    U_n(t)
    =
    \mathcal{T}
    \exp\left[
    -i\int_0^t H_n(t')\,dt'
    \right].
\end{equation}
For constant detuning \(\delta\), the result is explicit. Defining
\begin{equation}
    \lambda_n
    =
    \sqrt{g^2 n+\frac{\delta^2}{4}},
\end{equation}
one obtains
\begin{equation}
    A_n(t)
    =
    \cos(\lambda_n t)
    +
    i\frac{\delta}{2\lambda_n}\sin(\lambda_n t),
    \label{eq:sm_A_n}
\end{equation}
and
\begin{equation}
    B_n(t)
    =
    -i\frac{g\sqrt n}{\lambda_n}\sin(\lambda_n t).
    \label{eq:sm_B_n}
\end{equation}
On resonance, \(\delta=0\), this reduces to
\begin{equation}
    A_n(t)=\cos(g\sqrt n\,t),
    \qquad
    B_n(t)=-i\sin(g\sqrt n\,t).
\end{equation}
Tracing out the battery gives the reduced qubit state. Writing
\begin{equation}
    \lvert \Psi(t)\rangle
    =
    \sum_{k=0}^{\infty}
    \left[
    G_k(t)\lvert k,g\rangle
    +
    E_k(t)\lvert k,e\rangle
    \right],
\end{equation}
with
\begin{equation}
    G_k(t)=c_k A_k(t),
    \qquad
    E_k(t)=c_{k+1}B_{k+1}(t),
\end{equation}
we find
\begin{equation}
    \rho_{ee}(t)
    =
    \sum_{n=1}^{\infty}
    |c_n|^2 |B_n(t)|^2,
    \label{eq:sm_rhoee_general}
\end{equation}
and
\begin{equation}
    \rho_{gg}(t)
    =
    \sum_{n=0}^{\infty}
    |c_n|^2 |A_n(t)|^2.
\end{equation}
The qubit coherence is
\begin{equation}
    \rho_{ge}(t)
    = \sum_{k=0}^{\infty}
    c_k c_{k+1}^*
    A_k(t)B_{k+1}^*(t).
    \label{eq:sm_qubit_coherence_adjacent}
\end{equation}
Equation~\eqref{eq:sm_qubit_coherence_adjacent} is central. A local qubit phase reference generated by the battery requires coherences between neighboring photon-number states, \(c_k c_{k+1}^*\). These are precisely the coherences that also determine
\begin{equation}
    \langle a_{\rm b}\rangle
    =
    \sum_{k=0}^{\infty}
    \sqrt{k+1}\,c_k^*c_{k+1}.
    \label{eq:sm_amean_number_coherences}
\end{equation}
Therefore a battery state may contain energy and still fail to provide a useful phase reference if it lacks adjacent number coherences.

\suppsubsection{Displaced squeezed battery state}

We use the displaced squeezed state
\begin{equation}
    \lvert \alpha,\zeta\rangle
    =
    D(\alpha)S(\zeta)\lvert 0\rangle,
    \label{eq:sm_displaced_squeezed_state}
\end{equation}
with
\begin{equation}
    D(\alpha)
    =
    \exp\left(\alpha a_{\rm b}^\dagger-\alpha^*a_{\rm b}\right),
\end{equation}
and
\begin{equation}
    S(\zeta)
    =
    \exp\left[
    \frac{1}{2}
    \left(
    \zeta^* a_{\rm b}^2-\zeta a_{\rm b}^{\dagger 2}
    \right)
    \right],
    \qquad
    \zeta=r e^{i\vartheta_s}.
\end{equation}
The mean photon number is
\begin{equation}
    \bar n
    = \langle a_{\rm b}^\dagger a_{\rm b}\rangle
    = |\alpha|^2+\sinh^2 r.
    \label{eq:sm_squeezed_nbar}
\end{equation}
Thus, when we compare different battery states at fixed total energy \(\bar n\), the displacement amplitude must be chosen as
\begin{equation}
    |\alpha|^2
    =
    \bar n-\sinh^2 r.
    \label{eq:sm_fixed_energy_alpha}
\end{equation}
This condition requires
\begin{equation}
    \bar n>\sinh^2 r.
\end{equation}

The coherent phase-reference fraction is defined as
\begin{equation}
    \eta_{\rm coh}
    =
    \frac{|\langle a_{\rm b}\rangle|^2}{\langle a_{\rm b}^\dagger a_{\rm b}\rangle}.
    \label{eq:sm_eta_def}
\end{equation}
For the displaced squeezed state,
\begin{equation}
    \langle a_{\rm b}\rangle=\alpha,
\end{equation}
so at fixed \(\bar n\),
\begin{equation}
    \eta_{\rm coh}
    =
    \frac{|\alpha|^2}{\bar n}
    =
    1-\frac{\sinh^2 r}{\bar n}.
    \label{eq:sm_eta_squeezed}
\end{equation}
Equation~\eqref{eq:sm_eta_squeezed} shows the basic tradeoff: increasing \(r\) can reduce number fluctuations, but it also transfers part of the fixed energy budget from coherent displacement into squeezed fluctuations. The latter do not provide a first-order phase reference for the linear Jaynes--Cummings coupling.

For a general squeezed angle, the photon-number variance of \(D(\alpha)S(\zeta)\lvert0\rangle\) is
\begin{equation}
    \operatorname{Var}(n)
    =
    |\alpha|^2
    \left[
    \cosh(2r)
    -
    \sinh(2r)\cos\left(2\varphi_\alpha-\vartheta_s\right)
    \right]
    +
    2\sinh^2 r\left(\sinh^2 r+1\right),
    \label{eq:sm_var_general_squeezed}
\end{equation}
where \(\alpha=|\alpha|e^{i\varphi_\alpha}\). The amplitude-squeezed choice aligns the squeezed quadrature with the displacement direction,
\begin{equation}
    \vartheta_s=2\varphi_\alpha.
\end{equation}
Then Eq.~\eqref{eq:sm_var_general_squeezed} reduces to
\begin{equation}
    \operatorname{Var}_{\rm amp}(n)
    =
    |\alpha|^2 e^{-2r}
    +
    2\sinh^2 r\left(\sinh^2 r+1\right).
    \label{eq:sm_var_amp_squeezed}
\end{equation}
Using Eq.~\eqref{eq:sm_fixed_energy_alpha}, this becomes
\begin{equation}
    \operatorname{Var}_{\rm amp}(n)
    =
    \left(\bar n-\sinh^2 r\right)e^{-2r}
    +
    2\sinh^2 r\left(\sinh^2 r+1\right).
    \label{eq:sm_var_amp_fixed_energy}
\end{equation}
For small \(r\),
\begin{equation}
    \operatorname{Var}_{\rm amp}(n)
    \simeq
    \bar n
    -
    2\bar n r
    +
    (2\bar n+1)r^2
    +
    O(r^3).
    \label{eq:sm_small_r_var}
\end{equation}
By contrast, the coherent state has
\begin{equation}
    \operatorname{Var}_{\rm coh}(n)=\bar n,
    \qquad
    \eta_{\rm coh}=1.
\end{equation}
Thus weak amplitude squeezing reduces photon-number fluctuations already at first order in \(r\), while the loss of coherent fraction begins as
\begin{equation}
    \eta_{\rm coh}
    =
    1-\frac{r^2}{\bar n}
    +
    O(r^4).
    \label{eq:sm_small_r_eta}
\end{equation}
This suggests that weak or moderate amplitude squeezing can be beneficial: it can narrow the spread of \(\Omega_n\) without strongly degrading the coherent phase reference.

The phase-squeezed choice corresponds to
\begin{equation}
    \vartheta_s=2\varphi_\alpha+\pi.
\end{equation}
In that case,
\begin{equation}
    \operatorname{Var}_{\rm phase}(n)
    =
    |\alpha|^2 e^{2r}
    +
    2\sinh^2 r\left(\sinh^2 r+1\right),
\end{equation}
which increases the photon-number spread and is therefore not expected to improve the finite-battery Landau--Zener contrast.

\suppsubsection{Gaussian number-squeezed phase states}
\begin{figure*}
    \centering
    \includegraphics[width=\textwidth]{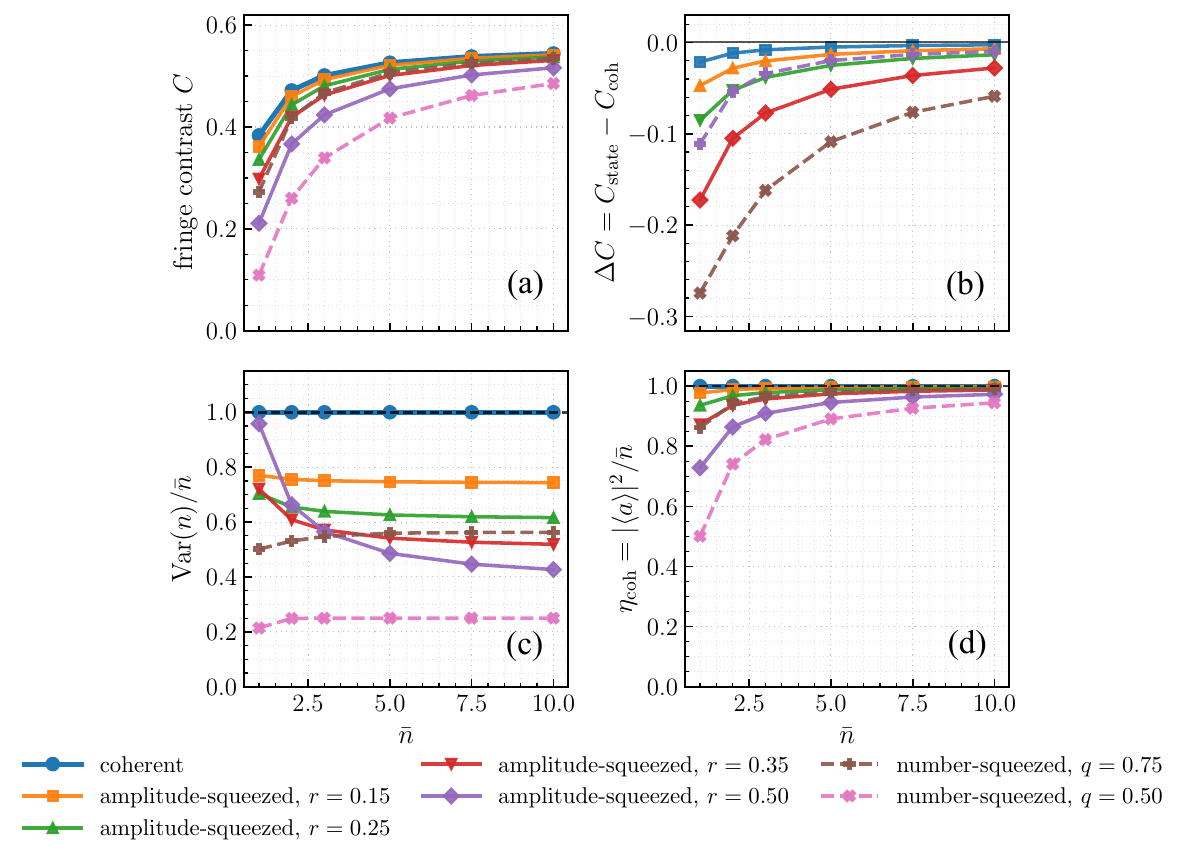}
    \caption{
    Squeezed-battery benchmark at fixed mean photon number. 
    (a) Fringe contrast \(C=\max_{\theta_{\rm geo}}P_e-\min_{\theta_{\rm geo}}P_e\) versus \(\bar n\) for a coherent battery, amplitude-squeezed coherent batteries with \(r=0.15,0.25,0.35,0.50\), and number-squeezed phase states. The coherent battery gives the largest contrast throughout the scanned range. 
    (b) Contrast change \(\Delta C=C_{\rm state}-C_{\rm coh}\), showing that all squeezed and number-squeezed states underperform the coherent state at the same total energy. 
    (c) Initial photon-number variance \({\rm Var}(n)\), showing that amplitude squeezing and number squeezing do narrow the distribution of Landau--Zener gaps \(\Omega_n=\Omega\sqrt{n/\bar n}\). 
    (d) Phase-reference fraction \(\eta_{\rm coh}=|\langle a_{\rm b}\rangle|^2/\bar n\), showing the cost of squeezing at fixed energy. The reduced number fluctuations in panel (c) are outweighed by the loss of coherent phase-reference strength in panel (d), explaining the contrast reduction in panels (a,b).\justifying}
    \label{fig:sm_squeezing_benchmark}
\end{figure*}

In addition to displaced squeezed states, Fig.~\ref{fig:sm_squeezing_benchmark}
also uses a family of number-squeezed phase states. These states are not meant
to represent a unique preparation protocol. They provide a clean diagnostic
family in which the photon-number width can be reduced while a controlled
phase relation between neighboring Fock components is retained.
For a target mean photon number \(\bar n\), we define
\begin{equation}
    |\psi_{q,\phi}\rangle
    =
    \sum_{n=0}^{N_{\rm cut}-1}
    c_n^{(q,\phi)} |n\rangle ,
    \label{eq:SM_number_squeezed_state}
\end{equation}
with coefficients
\begin{equation}
    c_n^{(q,\phi)}
    =
    \sqrt{p_n^{(q)}}\,e^{-in\phi}.
    \label{eq:SM_number_squeezed_coefficients}
\end{equation}
Here the phase \(\phi\) is the same battery phase used for the coherent state.
In the main text notation, \(\phi=\phi_\theta=\theta_{\rm geo}-\pi/2\). Thus
the adjacent Fock components have a fixed phase difference \(e^{-i\phi}\),
so the state carries a first-order phase reference whenever neighboring
probabilities overlap.
The photon-number probabilities are chosen as a truncated discrete Gaussian,
\begin{equation}
    p_n^{(q)}
    =
    \frac{1}{Z_q}
    \exp\left[
        -\frac{(n-\mu_q)^2}{2\sigma_q^2}
    \right],
    \qquad
    n=0,\ldots,N_{\rm cut}-1 ,
    \label{eq:SM_number_squeezed_probabilities}
\end{equation}
where
\begin{equation}
    Z_q
    =
    \sum_{n=0}^{N_{\rm cut}-1}
    \exp\left[
        -\frac{(n-\mu_q)^2}{2\sigma_q^2}
    \right]
    \label{eq:SM_number_squeezed_normalization}
\end{equation}
normalizes the state. The width parameter is
\begin{equation}
    \sigma_q = q\sqrt{\bar n}.
    \label{eq:SM_number_squeezed_width}
\end{equation}
Thus \(q=1\) corresponds to a Poisson-like width on the scale of a coherent
state, while \(q<1\) gives a number-squeezed distribution. In Fig.~\ref{fig:sm_squeezing_benchmark}
we use \(q=0.75\) and \(q=0.50\).
The center \(\mu_q\) is not fixed a priori to \(\bar n\). Instead, it is chosen
so that the truncated distribution has the desired mean energy,
\begin{equation}
    \sum_{n=0}^{N_{\rm cut}-1}
    n\,p_n^{(q)}
    =
    \bar n .
    \label{eq:SM_number_squeezed_mean_matching}
\end{equation}
This adjustment removes small shifts caused by the lower boundary at \(n=0\)
and by the finite numerical cutoff. With this convention,
\begin{equation}
    \langle a_{\rm b}^\dagger a_{\rm b}\rangle=\bar n
\end{equation}
by construction, and the photon-number variance is
\begin{equation}
    {\rm Var}_q(n)
    =
    \sum_{n=0}^{N_{\rm cut}-1}
    n^2 p_n^{(q)}
    -
    \bar n^2 .
    \label{eq:SM_number_squeezed_variance}
\end{equation}
For distributions narrow compared with the cutoff and sufficiently far from
the \(n=0\) boundary, this variance is approximately
\({\rm Var}_q(n)\simeq q^2\bar n\). The exact value plotted in
Fig.~\ref{fig:sm_squeezing_benchmark} is always computed from
Eq.~\eqref{eq:SM_number_squeezed_variance}.
The coherent phase-reference amplitude is
\begin{equation}
    \langle a_{\rm b}\rangle
    =
    \sum_{n=0}^{N_{\rm cut}-2}
    \sqrt{n+1}\,
    \left(c_n^{(q,\phi)}\right)^*
    c_{n+1}^{(q,\phi)} .
    \label{eq:SM_number_squeezed_a_mean_definition}
\end{equation}
Using Eq.~\eqref{eq:SM_number_squeezed_coefficients}, this becomes
\begin{equation}
    \langle a_{\rm b}\rangle
    =
    e^{-i\phi}
    \sum_{n=0}^{N_{\rm cut}-2}
    \sqrt{n+1}\,
    \sqrt{p_n^{(q)}p_{n+1}^{(q)}} .
    \label{eq:SM_number_squeezed_a_mean}
\end{equation}
Therefore the magnitude of the phase reference is controlled by the overlap
between neighboring photon-number probabilities. The corresponding coherent
phase-reference fraction is
\begin{equation}
    \eta_{\rm coh}^{(q)}
    =
    \frac{|\langle a_{\rm b}\rangle|^2}{\bar n}
    =
    \frac{1}{\bar n}
    \left(
        \sum_{n=0}^{N_{\rm cut}-2}
        \sqrt{n+1}\,
        \sqrt{p_n^{(q)}p_{n+1}^{(q)}}
    \right)^2 .
    \label{eq:SM_number_squeezed_eta}
\end{equation}
Equation~\eqref{eq:SM_number_squeezed_eta} shows explicitly why strong
number squeezing can be detrimental for geometric control. Reducing \(q\)
narrows the photon-number distribution and therefore reduces the spread of
sector-dependent gaps \(\Omega_n=2g\sqrt n\). At the same time, it reduces
the overlap between neighboring Fock components and hence reduces
\(|\langle a_{\rm b}\rangle|^2/\bar n\), the first-order phase reference that fixes
the transverse control axis. In the limiting case of an exact Fock state,
there is no neighboring-number overlap, \(\langle a_{\rm b}\rangle=0\), and therefore
\(\eta_{\rm coh}=0\), even though the stored energy is sharply defined.

\suppsubsection{Squeezed vacuum as a limiting case}

The squeezed vacuum is
\begin{equation}
    \lvert \zeta\rangle
    =
    S(\zeta)\lvert 0\rangle,
    \qquad
    \zeta=re^{i\vartheta_s}.
\end{equation}
Its Fock expansion is
\begin{equation}
    \lvert \zeta\rangle
    =
    \frac{1}{\sqrt{\cosh r}}
    \sum_{m=0}^{\infty}
    \left(
    -e^{i\vartheta_s}\tanh r
    \right)^m
    \frac{\sqrt{(2m)!}}{2^m m!}
    \lvert 2m\rangle.
    \label{eq:sm_squeezed_vacuum_expansion}
\end{equation}
Only even photon numbers are present. Hence
\begin{equation}
    c_k c_{k+1}^*=0
    \qquad
    \text{for all }k,
\end{equation}
and therefore
\begin{equation}
    \langle a_{\rm b}\rangle=0,
    \qquad
    \eta_{\rm coh}=0,
    \qquad
    \rho_{ge}(t)=0.
\end{equation}
The squeezed vacuum can still exchange energy with the qubit because it has nonzero photon number,
\begin{equation}
    \langle a_{\rm b}^\dagger a_{\rm b}\rangle=\sinh^2 r,
\end{equation}
but it does not supply a first-order phase reference for the linear Jaynes--Cummings coupling. For example, on resonance,
\begin{equation}
    P_e(t)
    =
    \sum_{m=1}^{\infty}
    p_{2m}
    \sin^2(g\sqrt{2m}\,t),
\end{equation}
where
\begin{equation}
    p_{2m}
    =
    \frac{(2m)!}{2^{2m}(m!)^2}
    \frac{\tanh^{2m}r}{\cosh r}.
\end{equation}
This population depends on the photon-number distribution, but not on the squeezing angle \(\vartheta_s\). The squeezed vacuum therefore illustrates why energy alone is insufficient for geometric Landau--Zener interferometry: a state can contain battery energy but fail to define the azimuthal phase of the transverse qubit coupling.

\suppsubsection{Implications for the finite-battery interferometer}

The displaced amplitude-squeezed state can improve the interferometer only through a competition between two effects. The beneficial effect is the reduction of photon-number fluctuations,
\begin{equation}
    \operatorname{Var}(n)
    <
    \bar n,
\end{equation}
which narrows the distribution of sector-dependent gaps \(\Omega_n\). From Eq.~\eqref{eq:sm_gap_variance}, this suppresses the inhomogeneous broadening of the Landau--Zener beam splitters.

The detrimental effect is the reduction of the coherent phase-reference fraction,
\begin{equation}
    \eta_{\rm coh}
    =
    1-\frac{\sinh^2 r}{\bar n}.
\end{equation}
In the large-\(\bar n\) correspondence limit, the coherent displacement produces an effective transverse field with amplitude
\begin{equation}
    \Omega_{\rm eff}
    =
    2g|\langle a_{\rm b}\rangle|
    =
    2g|\alpha|.
\end{equation}
With the calibration \(g=\Omega/(2\sqrt{\bar n})\), this becomes
\begin{equation}
    \Omega_{\rm eff}
    =
    \Omega\sqrt{\eta_{\rm coh}}.
    \label{eq:sm_omega_eff_eta}
\end{equation}
Thus a squeezed state at fixed total energy generally has a smaller classical-like drive component than a coherent state with the same \(\bar n\).

The coherent state occupies a special point in this tradeoff. It has
\begin{equation}
    |\langle a_{\rm b}\rangle|^2=\bar n,
    \qquad
    \eta_{\rm coh}=1,
    \qquad
    \operatorname{Var}(n)=\bar n.
\end{equation}
Amplitude squeezing can reduce \(\operatorname{Var}(n)\), but only by sacrificing part of the coherent displacement energy. Consequently, a squeezed coherent battery can outperform the coherent battery only in a parameter regime where the improvement from reduced gap broadening is larger than the loss of phase-reference strength. This is not guaranteed and must be checked with the full qubit--battery dynamics.

\suppsubsection{Numerical implementation}

In the numerical calculations we do not replace the battery operator by a mean field. The qubit--battery density matrix is evolved with the full operator \(a_{\rm b}\), as in Eq.~\eqref{eq:sm_jc_hamiltonian}. The coupling is calibrated as
\begin{equation}
    g=\frac{\Omega}{2\sqrt{\bar n}},
\end{equation}
so that the central photon-number sector has the same nominal transverse gap as the classical reference protocol. The plotted geometric control parameter is implemented through the battery phase
\begin{equation}
    \phi_{\rm batt}=\theta_{\rm geo}-\frac{\pi}{2}.
\end{equation}
For a coherent battery we use
\begin{equation}
    \alpha=\sqrt{\bar n}\,e^{-i\phi_{\rm batt}}.
\end{equation}
For a displaced squeezed battery at fixed total energy, the coherent displacement is instead chosen as
\begin{equation}
    \alpha
    =
    \sqrt{\bar n-\sinh^2 r}\,
    e^{-i\phi_{\rm batt}},
    \label{eq:sm_alpha_numerical}
\end{equation}
and the amplitude-squeezed choice is
\begin{equation}
    \zeta
    =
    r e^{-2i\phi_{\rm batt}}.
    \label{eq:sm_zeta_numerical}
\end{equation}
Equations~\eqref{eq:sm_alpha_numerical} and \eqref{eq:sm_zeta_numerical} align the squeezed quadrature with the coherent displacement. The squeezing values used in the benchmark are
\begin{equation}
    r=0.15,\ 0.25,\ 0.35,\ 0.50.
\end{equation}
The value \(r=0.35\) is used as a representative moderate-squeezing case. It visibly reduces number fluctuations while still satisfying \(\sinh^2 r\ll \bar n\) for the main values of \(\bar n\), so that the coherent phase-reference fraction remains close to unity. Larger values of \(r\) are useful as stress tests, but they can degrade the contrast by moving too much of the fixed energy budget into squeezed fluctuations rather than into coherent displacement.

The master equation is solved in Liouville space. We write the time-dependent Hamiltonian as
\begin{equation}
    H(t)=H_0+\delta(t)H_\delta ,
\end{equation}
and construct the corresponding sparse Liouvillian in the form
\begin{equation}
    \dot{\rho}_{\rm vec}(t)
    =
    \left[
        \mathcal{L}_0+\delta(t)\mathcal{L}_\delta
    \right]
    \rho_{\rm vec}(t),
\end{equation}
using column-stacking vectorization. The dissipative terms included in \(\mathcal{L}_0\) are the same as in the main text: qubit relaxation, pure dephasing, and weak battery loss. The evolution is performed with an adaptive eighth-order Runge--Kutta solver, using relative and absolute tolerances
\begin{equation}
    {\tt rtol}=2\times 10^{-7},
    \qquad
    {\tt atol}=2\times 10^{-9}.
\end{equation}
After each integration segment, the density matrix is symmetrized and renormalized to remove numerical roundoff errors.

The protocol is implemented piecewise. The first segment evolves the system during the forward sweep from \(-\delta_0\) to \(+\delta_0\). The plateau is split at \(t_m=\tau_C/2\), where the instantaneous echo operation is applied as
\begin{equation}
    \rho(t_m^+)
    =
    \left(I_{\rm b}\otimes U_\pi\right)
    \rho(t_m^-)
    \left(I_{\rm b}\otimes U_\pi^\dagger\right).
\end{equation}
The state is then evolved through the remaining plateau and the reverse sweep. This implementation therefore includes the echo-induced redistribution between neighboring Jaynes--Cummings excitation sectors exactly.

The bosonic Hilbert space is truncated at a state-dependent cutoff \(N_{\rm cut}\). For coherent states we use
\begin{equation}
    N_{\rm cut}^{\rm coh}
    =
    \max\left[
        8,\,
        \left\lceil
            \bar n+5\sqrt{\bar n+1}+8
        \right\rceil
    \right].
\end{equation}
For displaced squeezed states the cutoff is enlarged conservatively by replacing \(\bar n\) in this estimate by
\begin{equation}
    \bar n_{\rm eff}
    =
    \bar n+4\sinh^2 r+2
\end{equation}
and using seven standard deviations instead of five, with a minimum cutoff of \(12\). For the Gaussian number-squeezed phase states, with width
\begin{equation}
    \sigma_n=\max\left(0.2,\ q\sqrt{\bar n}\right),
\end{equation}
we use
\begin{equation}
    N_{\rm cut}^{(q)}
    =
    \max\left[
        10,\,
        \left\lceil
            \bar n+8\sigma_n+10
        \right\rceil
    \right].
\end{equation}
All truncated battery states are normalized after construction. For the number-squeezed phase states, the center of the truncated Gaussian distribution is chosen numerically so that the initial mean photon number is exactly \(\bar n\) within the truncated Hilbert space.

The phase fringes and contrast benchmarks are computed on a uniform grid of \(101\) values of \(\theta_{\rm geo}\in[0,2\pi]\). The heatmaps use a uniform \(101\times 101\) grid in \((\theta_{\rm geo},\tau_p)\), with \(\theta_{\rm geo}\in[0,2\pi]\) and \(\tau_p\in[25,35]\,{\rm ns}\). The contrast is extracted directly from the numerical fringe as
\begin{equation}
    C
    =
    \max_{\theta_{\rm geo}} P_e(\theta_{\rm geo})
    -
    \min_{\theta_{\rm geo}} P_e(\theta_{\rm geo}).
\end{equation}
The coherent quantum-to-classical scaling is evaluated for
\begin{equation}
    \bar n
    =
    0.5,\ 0.8,\ 1.0,\ 1.5,\ 2.0,\ 3.0,\ 5.0,\ 7.5,\ 10.0,\ 15.0,
\end{equation}
while the squeezed-state benchmark in Fig.~\ref{fig:sm_squeezing_benchmark} uses
\begin{equation}
    \bar n
    =
    1.0,\ 2.0,\ 3.0,\ 5.0,\ 7.5,\ 10.0.
\end{equation}
For each initial battery state we record the initial photon-number variance, the coherent displacement strength, and the coherent phase-reference fraction,
\begin{equation}
    \operatorname{Var}(n),
    \qquad
    |\langle a_{\rm b}\rangle|^2,
    \qquad
    \eta_{\rm coh}
    =
    \frac{|\langle a_{\rm b}\rangle|^2}{\bar n}.
\end{equation}
These observables distinguish the two mechanisms responsible for contrast changes: narrowing of the photon-number distribution and loss of coherent phase-reference strength.

As shown in Fig.~\ref{fig:sm_squeezing_benchmark}, squeezed states reduce photon-number fluctuations and therefore narrow the distribution of sector-dependent Landau--Zener gaps. However, at fixed total battery energy this reduction comes at the cost of lowering \(\eta_{\rm coh}\). For the parameters considered here, the loss of phase-reference strength dominates over the benefit of reduced gap broadening, and the coherent battery yields the largest GLZI contrast. This shows that suppressing number fluctuations alone is not sufficient to improve a finite quantum-battery interferometer. The battery must also provide a stable first-order phase reference. In the present regime, the coherent state gives the best balance: it has maximal \(\eta_{\rm coh}=1\), while its residual photon-number fluctuations are less detrimental than the loss of coherent displacement in the squeezed states.

\twocolumngrid  

\bibliography{References}

\end{document}